\documentclass[12pt,reqno]{amsart}
\pdfoutput=1 
\usepackage{amsmath,bbm}
\usepackage{latexsym}
\usepackage{amsfonts}
\usepackage{amssymb}
\usepackage{color}
\usepackage{graphicx}
\usepackage{url}
\usepackage{enumerate}
\usepackage{tikz}
\usetikzlibrary{shadings, intersections, calc, plotmarks}
\usepackage{marginnote}
\usepackage{stackrel}

\usepackage{geometry}
\xdefinecolor{tumblue}     {RGB}{0,101,189}
\xdefinecolor{tumgreen}    {RGB}{162,173,  0}
\xdefinecolor{tumred}      {RGB}{229, 52, 24}
\xdefinecolor{tumivory}    {RGB}{218,215,203}
\xdefinecolor{tumorange}   {RGB}{227,114, 34}
\xdefinecolor{tumlightblue}{RGB}{152,198,234}

\newtheorem{theorem}{Theorem}
\newtheorem*{theorem*}{Theorem}
\newtheorem{lemma}{Lemma}
\newtheorem{corollary}{Corollary}
\newtheorem*{corollary*}{Corollary}

\newtheorem{definition}{Definition}

\newcommand{\id}{{\rm{id}}} 

\newcommand{\cB}{{\mathcal{B}}}

\newcommand{\C}{\mathbbm{C}}

\newcommand{\cL}{\mathcal{L}}

\newcommand{\cT}{\mathcal{T}}

\newcommand{\cN}{\mathcal{N}}
\newcommand{\cA}{\mathcal{A}}
\newcommand{\cE}{\mathcal{E}}

\newcommand{\cM}{\mathcal{M}}
\newcommand{\1}{\mathbbm{1}}

\def\>{{\rangle}}
\def\<{{\langle}}
\def\Z{{\mathbb{Z}}}
\newcommand{\be}{\begin{equation}}
\newcommand{\ee}{\end{equation}}
\newcommand{\bea}{\begin{eqnarray}}
\newcommand{\eea}{\end{eqnarray}}

\newcommand{\ket}[1]{|#1\rangle} 
\newcommand{\bra}[1]{\langle#1|} 
\newcommand{\kb}[2]{|#1\rangle\!\langle#2|} 
\newcommand{\tr}[1]{\mathrm{tr}\left[#1\right]} 
\newcommand{\Tr}{\mathrm{tr}}
\newcommand{\norm}[1]{\left\lVert #1 \right\rVert}

\usepackage{amsaddr}

\makeatletter
\renewcommand{\@set@authors@addresses}{%
  \par
  \begingroup
  \centering
  \vspace{1em}

  {\normalfont\normalsize
    \author@andify\authors
    \authors\par
  }

  \medskip

  \def\author##1{}%
  \def\address##1##2{%
    \par\smallskip
    {\normalfont\small\ignorespaces##2\par}%
  }%
  \addresses
  \par
  \endgroup
}

\makeatletter
\renewcommand{\@settitle}{%
  \begin{center}
    \normalfont\LARGE\bfseries
    \@title\par
  \end{center}
}
\makeatother

\begin{document}

\title[Key Repeater Rate]{Secret key cannot be repeated through jointly undistillable states}

\author[Blakaj]{Vjosa Blakaj$^{1}$}
\email{vb@math.ku.dk}

\author[Theurer]{Thomas Theurer$^{1}$}
\email{tth@math.ku.dk}

\author[Horodecki]{Karol Horodecki$^{2}$}
\email{karol.horodecki@ug.edu.pl}

\address{%
$^{1}$Department of Mathematical Sciences,
University of Copenhagen,
Universitetsparken 5,
2100 Copenhagen {\O}, Denmark
}

\address{%
$^{2}$Institute of Informatics,
National Quantum Information Centre,
Faculty of Mathematics, Physics and Informatics,
University of Gda\'nsk,
Wita Stwosza 57,
80-308 Gda\'nsk, Poland
}

\begin{abstract} 
We prove that, under a joint-undistillability condition, it is impossible to approximate even a \textit{single} private bit arbitrarily well from any number of copies of a pair of finite-dimensional link states. As a corollary, we show that the quantum key repeater rate is zero for every pair of finite-dimensional PPT link states. Although each state may individually contain distillable secret key, arbitrary tripartite local operations and public classical communication cannot generate key between the endpoints that is secret from the repeater station. Entanglement may survive the repeater operation, but distillable secret key does not.
\end{abstract}

\maketitle

\section{Introduction}
\label{sec:introduction}

Establishing a quantum internet is a central objective of quantum information
science. One of its principal applications is to enable secure communication
between distant users by distributing secret keys through quantum
networks~\cite{Wehner2018}. Extending quantum communication over large
distances requires methods that overcome the loss and noise accumulated
during transmission. Quantum repeaters address this challenge by dividing
the communication line into shorter elementary links and processing the
quantum resources shared across them~\cite{Briegel1998}. In conventional
repeater architectures, entanglement purification improves the quality of
individual links, while entanglement swapping connects them: a joint
operation at an intermediate station establishes entanglement between
endpoints that need not have interacted directly. The resulting entanglement
can then be used to generate secret key between the distant users.

Secret-key generation, however, does not require distillable entanglement.
It has been shown that certain PPT
bound-entangled states have positive distillable secret key, even though
no maximally entangled pairs can be distilled from them by Local Operations
and Classical Communication (LOCC)~\cite{Horodecki2005,Horodecki2009}.  The existence of such states separates the resources needed for
entanglement distillation from those sufficient for quantum key
distribution (QKD), and raises the possibility of extending secure communication using links that cannot support conventional entanglement-distillation-based repeaters.

In ~\cite{Limitations} this question has been formulated in terms
of the quantum key repeater rate. Alice and Charlie share a
state $\rho_{AC_1}$, Charlie and Bob share a state $\sigma_{C_2B}$, and
the three parties use LOCC to generate key between Alice and Bob that is
secret from Charlie and the purifying environment. Can the secrecy in the
two links be extended through the intermediate station, as entanglement is
extended by entanglement swapping? Their work established upper bounds on
the repeater rate and exhibited families of PPT states whose distillable
key approaches one bit per copy while their repeater rate becomes
arbitrarily small. These results demonstrated a substantial separation
between the usefulness of a state for a single QKD link and its usefulness
in a repeater network. They left open whether PPT links can support a
strictly positive quantum key repeater rate.

In this work, by generalizing the inequality in~\cite[Lemma~3.2]{OpenAI}, we prove that from all pairs of link states $\rho_{AC_1},\sigma_{C_2B}$ that contain no distillable entanglement in the bipartition $AB:C_1C_2$, it is impossible to obtain a single approximation of a private state shared between Alice and Bob via tripartite separable operations, even if post-selection is allowed. This implies in particular that no key can be repeated through such states (which include pairs of PPT states). 
Thus, although in principle  entanglement may survive the repeater operation, the
resulting endpoint state has zero distillable key. Independent PPT
bound-entangled links therefore cannot support a positive quantum key
repeater rate in this model.

The methods we employ mirror the techniques used by Islam Faisal in Ref.~\cite{Faisal} to show that there exist bound entangled states with negative partial transpose.

\section{Preliminaries}
Throughout this work, we assume all Hilbert spaces to be finite dimensional. We consider three spatially separated parties: Alice, Bob, and Charlie. Their quantum systems and corresponding Hilbert spaces are denoted by \(A,A_1,A',\ldots\), \(B,B_1,B',\ldots\), and \(C,C_1,C',\ldots\), respectively.\vspace{0.1cm}

For a system $X$, we write $x=|X| := \dim X$ and fix a computational basis $\left\{\ket{i}_X\right\}_{i}^{|X|}$. We identify operators on $X$ with their matrices in this basis, and all transpositions are taken with respect to these fixed bases. We denote the algebra of $v \times v$  complex matrices by $M_v(\mathbb C)$ and the trace norm by 
\begin{equation}
    \|Z\|_1 := \tr{\sqrt{Z^\dagger Z}}.
\end{equation}

\begin{definition}
    [Distillable entanglement] The two-way distillable entanglement of a bipartite state \(\rho_{AB}\) is
\[
E_D^{A:B}(\rho_{AB})
:=
\inf_{\varepsilon>0}
\limsup_{n\to\infty}
\sup_{\substack{m\in\mathbb N_0\\
\Lambda_n\in\mathrm{LOCC}(A^n:B^n)}}
\left\{
\frac{m}{n}:
\Lambda_n\!\left(\rho_{AB}^{\otimes n}\right)
\approx_\varepsilon
\bigl(|\Psi\rangle\langle\Psi|\bigr)^{\otimes m}
\right\}.
\]
Here
\[
|\Psi\rangle=\frac{|00\rangle+|11\rangle}{\sqrt2},
\qquad
\omega\approx_\varepsilon\tau
\ \Longleftrightarrow\
\frac12\|\omega-\tau\|_1\leq\varepsilon,
\]
and the optimization is over trace-preserving LOCC maps allowing classical communication in both directions. The rate \(m/n\) counts the number of distilled maximally entangled qubit pairs per input copy.
    
\end{definition}

\begin{definition}[PPT states \cite{Peres1996,Horodecki1996}]
A bipartite density operator
$\rho_{AB}\in M_a(\mathbb C)\otimes M_b(\mathbb C)$,
has \emph{positive partial transpose}, or is a \emph{PPT state},
if
\[
\rho_{AB}^{T_B}
:=
(\operatorname{id}_A\otimes T_B)(\rho_{AB})
\geq 0,
\]
where $T_B$ denotes transposition on $B$. We denote the set of PPT states by $\mathrm{PPT}$ whenever
the bipartition is clear from the context.
\end{definition}

The target states in a quantum key repeater protocol are private states corresponding to an ideal cryptographically secret key shared between the endpoints. The set of private states was characterized in Refs.~\cite{Horodecki2005,Horodecki2009}.

\begin{definition}\label{def:privatestates}[Private states~\cite{Horodecki2005,Horodecki2009}]
Let $A = A_1 \otimes A_2$ and $B = B_1 \otimes B_2$, with $|A_1|=|B_1|=d$. A state $\gamma^{(d)}_{AB}$ is a \emph{private state} of key dimension $d$, or a \emph{pdit}, if it can be written as 
\begin{align}
        \gamma^{(d)}_{AB}=\frac{1}{d} \sum_{i,j=1}^{d}\kb{ii}{jj}_{A_1B_1}\otimes U_i \sigma_{A_2B_2} U_j^\dagger,
    \end{align}
where $\sigma_{A_2B_2}$ is a density operator and the $U_i$ are unitaries acting on $A_2B_2$.
The systems $A_1$ and $B_1$ are the \emph{key systems}, and $A_2$ and $B_2$ are the \emph{shield systems}. A private state with $d=2$ is called a \emph{private bit}, or a \emph{pbit}.
\end{definition}

We now define the rate at which private states can be established between Alice and Bob using the two links through Charlie.

\begin{definition}[Quantum key repeater rate \cite{PhysRevLett.119.220506, Limitations}]
Let $\rho_{AC_1}$ and $\sigma_{C_2B}$ be bipartite states shared
between Alice and Charlie, and Charlie and Bob, respectively.
The \emph{quantum key repeater rate} is the optimal asymptotic
rate at which private states can be distilled between Alice
and Bob by tripartite LOCC, with Charlie's systems traced
out at the end of the protocol. Formally,
\[
\begin{aligned}
K_{A\leftrightarrow C\leftrightarrow B}(\rho,\sigma)
:= {}&
\inf_{\varepsilon > 0}
\limsup_{n\to\infty}
\sup_{\mathcal E_n,\,\gamma_{A'B'}^{(d)}}
\\
&\left\{
\frac{\log_2 d}{n} : \frac{1}{2} \norm{
\operatorname{Tr}_{\mathrm{Charlie}}\left[
\mathcal E_n
\!\left(
(\rho_{AC_1}\otimes\sigma_{C_2B})^{\otimes n}
\right)\right]-
\gamma_{A'B'}^{(d)}}_1\le \varepsilon
\right\}.
\end{aligned}
\]
Here, the supremum is taken over trace-preserving tripartite
LOCC maps $\mathcal E_n$, acting with respect to the partition
\[
A^{\otimes n}:(C_1C_2)^{\otimes n}:B^{\otimes n},
\]
and all private states $\gamma_{A'B'}^{(d)}$ of key dimension
$d\geq 1$, with arbitrary finite-dimensional shield systems.

\end{definition}

\section{A Technical Lemma}
Recall that a linear map $\cN:M_r(\C)\longrightarrow M_a(\C)$ is
\emph{two-positive} if
\begin{align}\label{eq:two-positive-definition}
    (\id_2\otimes\cN)(Z)\geq0
    \qquad\text{for every }0\leq Z\in M_2(\C)\otimes M_r(\C),
\end{align}
where $\id$ denotes the identity map.
We generalize \cite[Lemma~3.2]{OpenAI} in two ways. Since the reversed block estimate in \cite[Lemma~3.2]{OpenAI} requires only two-positivity, we extend the result
beyond PPT maps. We also remove the restriction to rank-one operators.

Before stating the generalization, we define generalized represented operators.

\begin{definition}[Generalized represented operators]\label{def:gen_rep_op} A bipartite operator
	$$\omega_{AB} \in M_a(\mathbb{C})\otimes M_b(\mathbb{C})$$
	is a \emph{generalized represented operators} if there exists a positive semidefinite operator $\tau\in M_r(\C)\otimes M_s(\C)$, and completely positive maps
	$$
	\mathcal{L}:M_r(\mathbb{C})\longrightarrow M_a(\mathbb{C}),
	\qquad
	\mathcal{M}:M_s(\mathbb{C})\longrightarrow M_b(\mathbb{C}),
	$$
	such that 
	\begin{align}\label{eq:two-positive-assumption}
		(\cL\otimes\cM)\circ \cT
	\end{align}
	is two-positive, and
	$$
	\omega_{AB}
	=
	(\mathcal{L}\otimes\mathcal{M})
	\bigl(\tau\bigr).
	$$
\end{definition}
Note that in the above Definition, the traces of $\tau$ and $\omega_{AB}$ are not necessarily equal to one, and the maps $\cL$ and $\cM$ are not necessarily trace-preserving. 

\begin{lemma}[Homogeneous four-operator inequality]\label{lem:OpenAI_generalized} Let $\omega_{AB}$ be a generalized represented operator and let
\begin{align}\label{eq:four-operators}
    X_0,X_1\in M_a(\C),
    \qquad
    Y_0,Y_1\in M_b(\C),
\end{align}
be positive semidefinite matrices.
For 
\begin{align}\label{eq:Dij-cij}
    D_{ij}:=X_i\otimes Y_j,
    \qquad
    c_{ij}:=\tr{D_{ij}\omega_{AB}}\ \forall i,j\in\{0,1\}
\end{align}
it then holds that 
\begin{align}\label{eq:MainLemmaOpenAI}
    \norm{D_{00}^{1/2}\,\omega_{AB}\,D_{11}^{1/2}}_1
    \leq
    \sqrt{c_{00}c_{01}}
    +\sqrt{c_{10}c_{11}}
    +2\sqrt{c_{01}c_{10}}.
\end{align}
\end{lemma}

\begin{proof}
Using Definition~\ref{def:gen_rep_op}, write
\begin{align}
	\omega_{AB}=
	(\mathcal{L}\otimes\mathcal{M})
	\bigl(\tau\bigr).
\end{align} 
We first consider the case where $\tau=\kb{\chi}{\chi}$ has rank-one. Via the Hilbert--Schmidt adjoints, define
\begin{align}\label{eq:pullback-operators}
    E_i:=\cL^\dagger(X_i)\geq0,
    \qquad
    F_j:=\cM^\dagger(Y_j)\geq0.
\end{align}
Without loss of generality, we can assume that $E_0$ and $E_1$ commute: To see this, let $\Pi$ be the orthogonal
projection onto the range of $(E_0+E_1)$, and let $S$ equal
$(E_0+E_1)^{-1/2}$ on that range and the identity on its orthogonal
complement. Thus, $S$ is positive and hence invertible, even when $E_0+E_1=0$.
Define
\begin{align}\label{eq:coordinate-change}
    \cL'(Z):=\cL(SZS^\dagger),
    \qquad
    \ket{\chi'}:=(S^{-1}\otimes\1_s)\ket{\chi}
\end{align}
and note that
\begin{align}\label{eq:unchanged-output}
    (\cL'\otimes\cM)(\kb{\chi'}{\chi'})
    =(\cL\otimes\cM)(\kb{\chi}{\chi})
    =\omega_{AB}.
\end{align}
Moreover, with 
\begin{align}\label{eq:commuting-pullbacks}
    E_i':=(\cL')^\dagger(X_i)=S^\dagger E_iS,
\end{align}
it holds that 
\begin{align}
	E_0'+E_1'=\Pi,
	\qquad
	[E_0',E_1']=0,
\end{align}
where the last assertion follows because $E_i'=\Pi E_i'\Pi$. Complete
positivity of $\cL'$ follows directly from its definition. Moreover,
writing $\overline S$ for the entrywise complex conjugation, we have
\begin{align}\label{eq:two-positivity-preserved}
    \bigl((\cL'\otimes\cM)\circ \cT\bigr)(Z)
    =\bigl((\cL\otimes\cM)\circ \cT\bigr)
    \!\left((\overline S\otimes\1_s)Z
    (\overline S\otimes\1_s)^\dagger\right).
\end{align}
The map $\overline S\cdot \overline S^\dagger$ is completely positive, so its 
composition on the input side preserves two-positivity. This way, we found a representation of $\omega_{AB}$ with the desired property. 

For vectors $\ket{u},\ket{v}\in\C^r\otimes\C^s$,
write
\begin{align}\label{eq:vector-weights}
    c_{ij}(u)
    :=\bra{u}(E_i\otimes F_j)\ket{u}
    =\tr{
        D_{ij}(\cL\otimes\cM)(\kb{u}{u})}.
\end{align}
In particular, $c_{ij}=c_{ij}(\chi)$. Applying
$\id_2\otimes(\cL\otimes\cM)$ to the block matrix
\begin{align}\label{eq:direct-input-block}
    \begin{pmatrix}
        \kb{u}{u}&\kb{u}{v}\\
        \kb{v}{u}&\kb{v}{v}
    \end{pmatrix}=\begin{pmatrix}
        \ket{u}\\
        \ket{v}
    \end{pmatrix} (\bra{u},\bra{v}),
\end{align}
gives a positive matrix. Conjugating by
$\operatorname{diag}(D_{00}^{1/2},D_{11}^{1/2})$ and using
that according to \cite[Lemma~3.1]{OpenAI}, for $H\in M_x(\C)$, $J \in M_y(\C)$, and $Z\in\C^{x\times y}$
\begin{align}\label{eq:positive-block-estimate}
    \begin{pmatrix}H&Z\\Z^\dagger&J\end{pmatrix}\geq0
    \text{ implies }
    \norm{Z}_1\leq\sqrt{\tr{H}\,\tr{J}},
\end{align} 
yields that
\begin{align}\label{eq:direct-estimate}
    \norm{D_{00}^{1/2}
        (\cL\otimes\cM)(\kb{u}{v})D_{11}^{1/2}}_1
    \leq\sqrt{c_{00}(u)c_{11}(v)}.
\end{align}
Analogously, apply
$\id_2\otimes\bigl((\cL\otimes\cM)\circ \cT\bigr)$ to the  block matrix
\begin{align}\label{eq:reversed-input-block}
    \begin{pmatrix}
        \kb{\overline v}{\overline v}&
        \kb{\overline v}{\overline u}\\
        \kb{\overline u}{\overline v}&
        \kb{\overline u}{\overline u}
    \end{pmatrix}.
\end{align}
Because
\begin{align}\label{eq:transpose-identities}
    (\kb{\overline v}{\overline u})^T=\kb{u}{v},
    \qquad
    (\kb{\overline u}{\overline u})^T=\kb{u}{u},
\end{align}
and by using two-positivity, the outcome is positive semidefinite and given by
\begin{align}\label{eq:reversed-output-block}
    \begin{pmatrix}
        (\cL\otimes\cM)(\kb{v}{v})&
        (\cL\otimes\cM)(\kb{u}{v})\\
        (\cL\otimes\cM)(\kb{v}{u})&
        (\cL\otimes\cM)(\kb{u}{u})
    \end{pmatrix}.
\end{align}
Analogously, we thus conclude that
\begin{align}\label{eq:reversed-estimate}
    \norm{D_{00}^{1/2}
        (\cL\otimes\cM)(\kb{u}{v})D_{11}^{1/2}}_1
    \leq\sqrt{c_{00}(v)c_{11}(u)}.
\end{align}

Let $P$ be the orthogonal projector onto the positive part of $E_0-E_1$, and define
\begin{align}\label{eq:vector-splitting}
    \ket{v}:=(P\otimes\1_s)\ket{\chi},
    \qquad
    \ket{u}:=((\1_r-P)\otimes\1_s)\ket{\chi}.
\end{align}
Then, $|\chi\rangle=|v\rangle+|u\rangle$.
Since $E_0$ and $E_1$ commute, $P$ commutes with each $E_i$. Consequently, 
\begin{equation}
\begin{aligned}
\langle v|(E_i\otimes F_j)|u\rangle
&=
\langle\chi|
\bigl(PE_i(\1_r-P)\otimes F_j\bigr)
|\chi\rangle\\
&=0.
\end{aligned}
\end{equation}
The conjugate cross term also vanishes, so 
\begin{align}\label{eq:split-weights}
    c_{ij}=c_{ij}(v)+c_{ij}(u),
    \qquad
    c_{11}(v)\leq c_{01}(v),
    \qquad
    c_{00}(u)\leq c_{10}(u),
\end{align}
where the inequalities follow by combining the facts that $E_0\geq E_1$ on the range of $P$, $E_0\leq E_1$ on the kernel of $P$, together with $F_0,F_1\geq0$. 

Finally, expanding
\[
|\chi\rangle\langle\chi|
=
|v\rangle\langle v|
+|u\rangle\langle u|
+|v\rangle\langle u|
+|u\rangle\langle v|
\]
and combining the above results with the triangle inequality 
proves \eqref{eq:MainLemmaOpenAI} for rank-one $\tau$. 

Now we consider the case of general $\tau\geq0$. For an eigendecomposition $\tau=\sum_k\kb{\chi_k}{\chi_k}$, let
\begin{align}\label{eq:mixed-decomposition}
    \omega_{AB}^{(k)}:=&(\cL\otimes\cM)(\kb{\chi_k}{\chi_k}), \nonumber \\
    c_{ij}^{(k)}:=&\tr{D_{ij}\omega_{AB}^{(k)}},
\end{align}
such that
\begin{align}\label{eq:mixed-weights}
    \omega_{AB}=\sum_k\omega_{AB}^{(k)},
    \qquad
    c_{ij}=\sum_k c_{ij}^{(k)}.
\end{align}
Each $\omega_{AB}^{(k)}$ is positive semidefinite. Applying the triangle inequality, the previously established rank-one bound to each summand, and the Cauchy–Schwarz inequality, respectively, gives
\begin{align}\label{eq:mixed-triangle}
    \norm{D_{00}^{1/2}\omega_{AB}D_{11}^{1/2}}_1
    &\leq\sum_k
    \norm{D_{00}^{1/2}\omega_{AB}^{(k)}D_{11}^{1/2}}_1,\\\label{eq:mixed-rank-one-bound}
    &\leq\sum_k\left(
        \sqrt{c_{00}^{(k)}c_{01}^{(k)}}
        +\sqrt{c_{10}^{(k)}c_{11}^{(k)}}
        +2\sqrt{c_{01}^{(k)}c_{10}^{(k)}}\right),\\\label{eq:mixed-cauchy-schwarz}
    &\leq
        \sqrt{c_{00}c_{01}}+\sqrt{c_{10}c_{11}}
        +2\sqrt{c_{01}c_{10}}.
\end{align}
\end{proof}

\section{Main results}

In this section, we establish a necessary condition for quantum key repetition. We show that if the joint link state has zero distillable entanglement across the bipartition \(AB:C_1C_2\), then every normalized endpoint state obtainable from any number of copies by a tripartite separable operation remains at trace distance at least \(1/10\) from every private bit, even when postselection is allowed. This implies that the quantum key repeater rate vanishes under the joint-undistillability assumption. The result for PPT links follows as a corollary.

\begin{theorem}\label{thm:workhorse}
Let the states $\rho_{AC_1}$ and $\sigma_{C_2B}$ be such that
\[
    E_D^{AB:C_1C_2}(\rho_{AC_1}\otimes\sigma_{C_2B})=0,
\]
and let $\gamma^{(2)}_{A'B'}$ be an arbitrary private bit. Moreover, let $n\ge 1$ and $\omega_{A'B'}$ be any state
attainable from  $(\rho_{AC_1}\otimes\sigma_{C_2B})^{\otimes n}$ by a
tripartite separable operation with nonzero probability. Then,
\begin{equation}\label{eq:workhorse-bound}
    \frac12\norm{\omega_{A'B'}-\gamma^{(2)}_{A'B'}}_1
    \geq\frac1{10}.
\end{equation}
\end{theorem}

\begin{proof}
Define the completely positive maps $\cL$ and $\cM$ by
\begin{align}
    \rho_{AC_1}^{\otimes n}
    &=\sum_{i,j}\cL(\kb{i}{j})_{A^{\otimes n}}
         \otimes\kb{i}{j}_{C_1^{\otimes n}},
         \label{eq:rhochoi}\\
    \sigma_{C_2B}^{\otimes n}
    &=\sum_{k,l}\kb{k}{l}_{C_2^{\otimes n}}
         \otimes\cM(\kb{k}{l})_{B^{\otimes n}}.
         \label{eq:sigmachoi}
\end{align}
The Choi operator of
$(\cL\otimes\cM)\circ\cT_{C_1^{\otimes n}C_2^{\otimes n}}$ is therefore
\[
    \bigl[(\rho_{AC_1}\otimes\sigma_{C_2B})^{\otimes n}\bigr]
        ^{T_{C_1^{\otimes n}C_2^{\otimes n}}}.
\]
By \cite[Theorem~4]{DiVincenzo2000} and the assumption of the Theorem, $(\cL\otimes\cM)\circ\cT_{C_1^{\otimes n}C_2^{\otimes n}}$ is thus two-positive. 

Consider a tripartite separable operation
\begin{equation}\label{eq:kraus}
    \cE(X)=\sum_r K_rXK_r^\dagger,
    \qquad
    K_r=\tilde A_r\otimes\tilde C_r\otimes\tilde B_r,
\end{equation}
where $\tilde C_r$ acts on $C_1^{\otimes n}C_2^{\otimes n}$. Define
\[
    \cA_r(X)=\tilde A_rX\tilde A_r^\dagger,
    \qquad
    \cB_r(X)=\tilde B_rX\tilde B_r^\dagger,
\]
and thus
\begin{align}\label{eq:branch}
    W_r :=&\Tr_{\rm Charlie}\bigl[
        K_r(\rho_{AC_1}\otimes\sigma_{C_2B})^{\otimes n}K_r^\dagger
      \bigr]\nonumber\\
      =&(\cA_r\otimes \cB_r)\ \Tr_{\rm Charlie}\bigl[(\mathbbm{1}_{A^{\otimes n}}\otimes \tilde C_r^\dagger \tilde C_r \otimes \mathbbm{1}_{B^{\otimes n}})(\rho_{AC_1}\otimes\sigma_{C_2B})^{\otimes n}\bigr]\notag\\
        =&(\cA_r\otimes \cB_r)\ \sum_{i,j,k,l}\langle j,l|\tilde C_r^\dagger \tilde C_r|i,k\rangle\,
        \cL(\kb{i}{j})\otimes\cM(\kb{k}{l})\notag \\
    =&\bigl((\cA_r\circ\cL)\otimes(\cB_r\circ\cM)\bigr)
      \bigl((\tilde C_r^\dagger\tilde C_r)^T\bigr).
\end{align}
Composition with the completely positive map $\cA_r\otimes\cB_r$ 
preserves two-positivity and thus each $W_r$ is a generalized represented operator. 

Now write $A'=A'_1\otimes A'_2$ and $B'=B'_1\otimes B'_2$, where
$A'_1$ and $B'_1$ are the key systems. Set
\[
    X_i=\kb{i}{i}_{A'_1}\otimes\1_{A'_2},
    \qquad Y_j=\kb{j}{j}_{B'_1}\otimes\1_{B'_2},
    \qquad P_{ij}=X_i\otimes Y_j.
\]
For $\omega_{A'B'}=(\sum_r W_r)/\tr{\sum_r W_r}$, put
\[
    p_{ij}=\tr{P_{ij}\omega_{A'B'}},
    \qquad e=p_{01}+p_{10},
    \qquad z=\norm{P_{00}\omega_{A'B'}P_{11}}_1.
\]
From the triangle inequality,  Lemma~\ref{lem:OpenAI_generalized} applied to each $W_r$, and Cauchy--Schwarz, it follows that
\begin{equation}\label{eq:coherenceconstraint}
    z\leq\sqrt{p_{00}p_{01}}+\sqrt{p_{10}p_{11}}
       +2\sqrt{p_{01}p_{10}}
    \leq\sqrt{e(1-e)}+e.
\end{equation} 

To finish the proof, we must show that
\begin{equation}\label{eq:bound_e}
    \frac12\|\omega_{A'B'}-\gamma^{(2)}_{A'B'}\|_1 \geq 1/10,
\end{equation}
which we will do by contradiction: Assume that this is not the case. Since $\gamma^{(2)}_{A'B'}$ is a private bit, its corresponding keys agree perfectly, i.e.,
\begin{equation}\label{eq:key_pbit}
\operatorname{tr}\!\left[(P_{01}+P_{10})\gamma^{(2)}_{A'B'}\right]=0.
\end{equation}
Moreover, Definition \ref{def:privatestates} and unitary invariance of the trace norm imply

\begin{equation}
    \left\|P_{00}\gamma^{(2)}_{A'B'}P_{11}\right\|_1=\frac12.
\end{equation}
Since \(0\leq P_{01}+P_{10}\leq \1 \), the variational characterization of trace distance, together with Eq.~\eqref{eq:key_pbit}, gives
\[
\begin{aligned}
e = p_{01}+p_{10}
&=\operatorname{tr}\!\left[
(P_{01}+P_{10})
\bigl(\omega_{A'B'}-\gamma^{(2)}_{A'B'}\bigr)
\right]\\
&\le
\frac12\left\|\omega_{A'B'}-\gamma^{(2)}_{A'B'}\right\|_1
<\frac1{10}.
\end{aligned}
\]
Even more, since $P_{00}P_{11}=0$, and $\omega_{A'B'}-\gamma^{(2)}_{A'B'}$ is a Hermitian matrix, we get
\[
\left\|
P_{00}\bigl(\omega_{A'B'}-\gamma^{(2)}_{A'B'}\bigr)P_{11}
\right\|_1
\le
\frac12\left\|\omega_{A'B'}-\gamma^{(2)}_{A'B'}\right\|_1.
\]
Now, by the reverse triangle inequality, 
\begin{equation}\label{eq:zbigger25}
\begin{aligned}
z
&=\|P_{00}\omega_{A'B'}P_{11}\|_1\\
&\ge
\left\|P_{00}\gamma^{(2)}_{A'B'}P_{11}\right\|_1
-
\left\|P_{00}
\bigl(\omega_{A'B'}-\gamma^{(2)}_{A'B'}\bigr)
P_{11}\right\|_1\\
&\ge
\frac12-
\frac12\left\|\omega_{A'B'}-\gamma^{(2)}_{A'B'}\right\|_1
>\frac25.
\end{aligned}
\end{equation}
However, inequality (\ref{eq:coherenceconstraint}) with $0 \leq e < 1/10$ gives that 
\begin{equation}
    z < 2/5,
\end{equation}
because $f(e) := \sqrt{e(1-e)} + e$ is increasing in $[0, 1/10]$, contradicting (\ref{eq:zbigger25}). Therefore,
\[
\frac12\left\|\omega_{A'B'}-\gamma^{(2)}_{A'B'}\right\|_1
\ge\frac1{10}.
\]

\end{proof}

The theorem shows that, for any pair of states $\rho_{AC_1}$ and $\sigma_{C_2B}$ satisfying its assumptions, no number of copies allows Alice and Bob to approximate even a \textit{single} private bit
arbitrarily well using tripartite separable operations, even with postselection. Since every LOCC operation is separable, the same conclusion holds for LOCC protocols. In particular, every pair of PPT link states satisfies the joint-undistillability assumption
of the theorem. We therefore obtain the following corollary.

\begin{corollary}\label{thm:main}
Let \(\rho_{AC_1}\) and \(\sigma_{C_2B}\) satisfy
\[
E_D^{AB:C_1C_2}
(\rho_{AC_1}\otimes\sigma_{C_2B})=0.
\]
Then
\[
K_{A\leftrightarrow C\leftrightarrow B}(\rho,\sigma)=0.
\]
In particular, this holds whenever both link states are PPT.
\end{corollary}

\vspace{0.3cm}

\noindent {\it Acknowledgments.} VB acknowledges support from VILLUM FONDEN via the QMATH Centre of Excellence (Grant No.~10059), and the Novo Nordisk Foundation (Grant No.~NNF20OC0059939, “Quantum for Life”). TT acknowledges support from a Postdoc Scholarship on Quantum Algorithms or Quantum Software from the Danish e-infrastructure Consortium (DeiC). KH acknowledges National Science Centre grant Opus 25 no. UMO-2023/49/B/ST2/02468. The authors acknowledge discussions with Matthias Christandl and Alexander Müller-Hermes. 
The authors used ChatGPT-6 Astra to assist during the preparation of this manuscript. The authors reviewed and revised the resulting material and remain responsible for the final manuscript.\vspace{0.5cm}

\noindent {\it Remark.}
After the completion of this work, the preprint “Arbitrarily large link key, zero repeater key: two PPT barriers in quantum networks” by Zixuan He \cite{PPT2}, was brought to the authors’ attention. The two works are completely independent.
\bibliographystyle{halpha}
\bibliography{bibliography}{}

@article{DiVincenzo2000,
  title = {Evidence for bound entangled states with negative partial transpose},
  author = {DiVincenzo, David P. and Shor, Peter W. and Smolin, John A. and Terhal, Barbara M. and Thapliyal, Ashish V.},
  journal = {Phys. Rev. A},
  volume = {61},
  issue = {6},
  pages = {062312},
  numpages = {13},
  year = {2000},
  month = {May},
  publisher = {American Physical Society},
  doi = {10.1103/PhysRevA.61.062312},
  url = {https://link.aps.org/doi/10.1103/PhysRevA.61.062312}
}

@article{PPT2,
  title = {Arbitrarily large link key, zero repeater key: two PPT barriers in quantum networks},
  author = {He, Zixuan},
  journal = {https://doi.org/10.5281/zenodo.23200366},
  volume = {},
  issue = {},
  pages = {},
  year = {2026},
  publisher = {},
  doi = {},
  url = {}
}

@article{Faisal,
  title = {Bound Entanglement with Negative Partial Transpose Exists},
  author = {Faisal, Islam},
  journal = {arXiv:2611.34921v1 },
  volume = {},
  issue = {},
  pages = {},
  year = {2026},
  publisher = {},
  doi = {},
  url = {}
}

@article{Wehner2018,
  title = {Quantum internet: A vision for the road ahead},
  author = {Wehner, Stephanie and Elkouss, David and Hanson, Ronald},
  journal = {Science},
  volume = {362},
  issue = {6412},
  pages = {eaam9288},
  year = {2018},
  publisher = {American Association for the Advancement of Science},
  doi = {10.1126/science.aam9288},
  url = {https://doi.org/10.1126/science.aam9288}
}

@article{Briegel1998,
  title = {Quantum Repeaters: The Role of Imperfect Local Operations in Quantum Communication},
  author = {Briegel, Hans-J{\"u}rgen and D{\"u}r, Wolfgang and Cirac, J. Ignacio and Zoller, Peter},
  journal = {Phys. Rev. Lett.},
  volume = {81},
  pages = {5932--5935},
  year = {1998},
  publisher = {American Physical Society},
  doi = {10.1103/PhysRevLett.81.5932},
  url = {https://link.aps.org/doi/10.1103/PhysRevLett.81.5932}
}

@article{Horodecki2005,
  title = {Secure Key from Bound Entanglement},
  author = {Horodecki, Karol and Horodecki, Micha{\l{}} and Horodecki, Pawe{\l{}} and Oppenheim, Jonathan},
  journal = {Phys. Rev. Lett.},
  volume = {94},
  issue = {16},
  pages = {160502},
  numpages = {4},
  year = {2005},
  month = {Apr},
  publisher = {American Physical Society},
  doi = {10.1103/PhysRevLett.94.160502},
  url = {https://link.aps.org/doi/10.1103/PhysRevLett.94.160502}
}

@article{Limitations,
  title = {Limitations on quantum key repeaters},
  author = {Bäuml, Stefan and Christandl, Matthias and Horodecki, Karol and Winter, Andreas},
  journal = {Nat Commun},
  volume = {6},
  issue = {6908},
  pages = {},
  numpages = {},
  year = {2015},
  month = {},
  publisher = {},
  doi = {},
  url = {https://doi.org/10.1038/ncomms7908}
}

@article{PhysRevLett.119.220506,
  title = {Private States, Quantum Data Hiding, and the Swapping of Perfect Secrecy},
  author = {Christandl, Matthias and Ferrara, Roberto},
  journal = {Phys. Rev. Lett.},
  volume = {119},
  issue = {22},
  pages = {220506},
  numpages = {6},
  year = {2017},
  month = {Nov},
  publisher = {American Physical Society},
  doi = {10.1103/PhysRevLett.119.220506},
  url = {https://link.aps.org/doi/10.1103/PhysRevLett.119.220506}
}

@ARTICLE{Horodecki2009,
  author={Horodecki, Karol and Horodecki, Michal and Horodecki, Pawel and Oppenheim, Jonathan},
  journal={IEEE Transactions on Information Theory}, 
  title={General Paradigm for Distilling Classical Key From Quantum States}, 
  year={2009},
  volume={55},
  number={4},
  pages={1898-1929},
  doi={10.1109/TIT.2008.2009798}}

@article{Peres1996,
  author  = {Peres, Asher},
  title   = {Separability Criterion for Density Matrices},
  journal = {Physical Review Letters},
  volume  = {77},
  number  = {8},
  pages   = {1413--1415},
  year    = {1996},
  doi     = {10.1103/PhysRevLett.77.1413}
}

@article{Horodecki1996,
  author  = {Horodecki, Micha{\l} and Horodecki, Pawe{\l}
             and Horodecki, Ryszard},
  title   = {Separability of Mixed States:
             Necessary and Sufficient Conditions},
  journal = {Physics Letters A},
  volume  = {223},
  number  = {1-2},
  pages   = {1--8},
  year    = {1996},
  doi     = {10.1016/S0375-9601(96)00706-2}
}

@article{OpenAI,
    author = {OpenAI},
    title = {Entanglement with zero distillable secret key in local dimension ten},
    journal = {Preprint dated 27 September 2026},
    year = {},
    doi = {https://github.com/openai/math/blob/main/preprints/Entanglement-with-zero-distillable-secret-key-in-local-dimension-ten-September-27-2026/paper.pdf}
}

\end{document}